\documentclass[preprintnumbers,prd,twocolumn,nofootinbib,superscriptaddress,nobibnotes,showpacs]{revtex4}
\usepackage{amsfonts}
\usepackage{mathrsfs}
\usepackage{epsfig}
\usepackage{graphicx}
\usepackage{dcolumn}
\usepackage{amsmath}

\begin{document}

\title{Probing light WIMPs with directional detection experiments}


\author{Ben Morgan}
%\email[]{ben.morgan@warwick.ac.uk}
%\thanks{}
\affiliation{Department of Physics, University of Warwick, Coventry, 
CV4 7AL, UK}
\author{Anne M. Green}
%\email[]{anne.green@nottingham.ac.uk}
%\thanks{}
\affiliation{School of Physics and Astronomy, University of 
  Nottingham, University Park, Nottingham, NG7 2RD, UK}


%\date{\today}

\begin{abstract}
The CoGeNT and CRESST WIMP direct detection experiments have recently
observed excesses of nuclear recoil events, while the DAMA/LIBRA 
experiment has a long standing annual modulation signal. 
It has been suggested that these excesses may be due to light mass, 
$m_{\chi}\sim5-10 \, {\rm GeV}$, WIMPs. 
The Earth's motion with respect to the Galactic rest frame leads to a 
directional dependence in the WIMP scattering rate, providing a powerful 
signal of the Galactic origin of any recoil excess. 
We investigate whether direct detection experiments with directional 
sensitivity have the potential to observe this anisotropic scattering rate 
with the elastically scattering light WIMPs proposed to explain the
observed excesses.
We find that the number of recoils required to detect an anisotropic
signal from light WIMPs at $5\sigma$ significance varies from 
7 to more than 190 over the set of target nuclei and energy thresholds 
expected for directional detectors. 
Smaller numbers arise from configurations where the detector is only
sensitive to recoils from the highest speed, and hence most anisotropic,
WIMPs. However, the event rate above threshold is very small in these cases,
leading to the need for large experimental exposures to accumulate even a
small number of events. To account for this sensitivity to the tail of the
WIMP velocity distribution, whose shape is not well known, we consider two
exemplar halo models spanning the range of possibilities. We also note that
for an accurate calculation the Earth's orbital speed must be averaged over.
We find that the exposures required to detect $10 \, {\rm GeV}$
WIMPs at a WIMP-proton cross-section of $10^{-4} \, {\rm pb}$ are of 
order $10^{3} \, {\rm kg \, day}$ for a $20 \, {\rm keV}$ 
energy threshold, within reach of planned directional detectors. Lower 
WIMP masses require higher exposures and/or lower energy thresholds for 
detection. 
\end{abstract}

%\keywords{}

\maketitle

%\section{}
\section{Introduction}

Direct detection experiments aim to detect dark matter in the form of
Weakly Interacting Massive Particles (WIMPs) via the nuclear recoils
which occur when WIMPs scatter off target nuclei~\cite{DD}. The
sensitivity of these experiments has increased rapidly over the last
few years, and they are probing the regions of WIMP mass-cross-section
parameter space populated by the lightest neutralino in Supersymmetric
extensions of the standard model (see e.g. Ref.~\cite{theory}).

Event rate excesses and annual modulations in various direct detection
experiments have prompted recent interest in light WIMPs. The DAMA
(now DAMA/LIBRA) collaboration have, for more than a decade, observed an
annual modulation of the event rate in their ${\rm NaI}$
crystals~\cite{dama}.  
 This annual modulation is consistent with light ($m_{\chi} \sim 5-10$ GeV) 
WIMPs scattering off ${\rm Na}$~\cite{Bottino:2003cz,gglightwimps}. The CoGeNT experiment, after allowing for 
backgrounds with an exponential plus constant energy spectrum, find an 
excess of low energy events which is consistent with WIMPs with mass 
$m_{\chi} \approx 7-11$ GeV~\cite{cogent1}. With a larger data set they 
have observed a 2.8 $\sigma$ annual modulation~\cite{cogent2}, with period 
and phase broadly consistent with the expectation for 
WIMPs~\cite{amtheory}. The CRESST experiment has observed an excess of 
events in their ${\rm CaWO}_{4}$ crystals above expectations from 
backgrounds~\cite{cresst}. The excess is compatible with either WIMPs of 
mass $m_{\chi} \sim 25$ GeV scattering off tungsten predominantly, or 
lighter, $m_{\chi} \sim 10$ GeV, WIMPs scattering off oxygen and calcium. 
It appears that it is not possible to explain all of these signals in 
terms of a single conventional elastic-scattering WIMP, especially when 
the exclusion limits from the CDMS~\cite{cdms}, XENON10~\cite{xenon10} 
and XENON100~\cite{xenon100} experiments and the CRESST commissioning 
data~\cite{cresstcomm} data are taken into account~\cite{ksz,khb,ox} 
(see also Refs.~\cite{hk,cogentamp}). None the less it is still possible 
that some subset of the putative signals are due to elastic scattering
light WIMPs. 

The deployment of a ${\rm NaI}$ detector at the South Pole has been 
proposed to directly test the DAMA annual modulation 
signal~\cite{damatest}. The direction dependence of the scattering 
rate~\cite{dirndep} provides another potentially powerful way of testing 
whether the observed excesses and annual modulations are due to elastic 
scattering light WIMPs. The amplitude of the directional signal is far 
larger than that of the annual modulation and hence the anisotropy of 
the WIMP induced nuclear recoils could be confirmed with a relatively 
small number of events~\cite{copi:krauss,pap1}. Furthermore the angular 
dependence of the recoils (in particular the peak recoil rate in the 
direction opposite to the motion of the solar system, or for low energy 
recoils  a ring around this direction~\cite{ring}) is extremely unlikely 
to be 
mimicked by backgrounds, and would allow unambiguous detection of
WIMPs~\cite{billard,pap5}. In this paper we investigate whether current 
and near future directional detectors would be able to detect elastic 
scattering light WIMPs.

\section{Modelling}
\label{model}
We use the same statistical techniques and methods for calculating the
directional nuclear recoil spectrum as in Refs.~\cite{pap1,pap2,pap3}.
We briefly summarise these procedures here, for further details see
these references. 

\subsection{Detector}
Most of the directional detectors currently under development 
(see Refs.~\cite{sm,cygnus} for reviews) are low pressure gas time 
projection chambers (TPCs), e.g. DMTPC~\cite{dmtpc}, DRIFT~\cite{drift},
MIMAC~\cite{mimac} and NEWAGE~\cite{newage}. Various gases have been 
considered, including ${\rm CF}_{4}$, ${\rm CS}_{2}$ and ${}^3{\rm He}$.
We therefore consider all four of these target nuclei: ${}^3{\rm He}$,
${\rm C}$, ${\rm F}$ and ${\rm S}$.  

Detailed calculations of the nuclear recoil track reconstruction are not 
available for all of these targets (see Ref.~\cite{billardtrack} for a 
detailed study of the reconstruction of simulated tracks for a MIMAC-like 
detector). Therefore we assume that the recoil directions are 
reconstructed perfectly in 3d. This is an optimistic assumption, therefore 
our results provide a lower limit on the number of events and exposure 
required by a real TPC based detector. Finite angular resolution does not 
significantly affect the number of events required to detect the 
anisotropic WIMP signal, provided it is not worse than of order tens of 
degrees~\cite{pap1,copi2d,billard11}. 2d read-out would, however, 
significantly degrade the detector capability~\cite{pap1,pap2,copi2d,pap3}.
We consider both vectorial and axial data i.e. where the senses of the 
recoils $+{\bf x}$ and $-{\bf x}$ are either measured for all recoil 
events or no events~\footnote{For studies of the effects of statistical 
sense determination see Refs.~\cite{pap4,billard11}.}. Sense 
discrimination is a major challenge for directional detectors. As discussed 
in detail in Ref.~\cite{billardtrack}, while the shape and charge 
distribution of nuclear recoil tracks are expected to be asymmetric, 
measuring these asymmetries with high efficiency is difficult in practice.

We consider four benchmark energy thresholds: $E_{\rm th}= 5, 10, 15$
and $20 \ {\rm keV}$ for each target.  Note that these are directional energy thresholds.
It is harder to measure the direction of a recoil than to simply detect it therefore, for a given experiment, the  
directional energy threshold is usually larger than the threshold for simply detecting recoils.
The high energy recoils are the most anisotropic~\cite{dirndep}, therefore
for heavy WIMPs a low energy threshold is not essential for directional 
detection. For instance for WIMPs with $m_{\chi} = 100 \, {\rm GeV}$ and 
a ${\rm S}$ target, reducing the energy threshold below $20 \, {\rm keV}$ 
does not significantly reduce the exposure required to reject
isotropy~\cite{pap3}. However for light WIMPs the differential event
rate decreases rapidly with increasing energy, and a low threshold is
crucial to obtain a non-negligible event rate. We discuss the viability of sufficiently low energy thresholds in Sec.~\ref{res}.

\subsection{WIMP masses and cross-sections}
We consider three benchmark light  WIMP masses, $m_{\chi} = 5, 7.5$ and 
$10 \, {\rm GeV}$ spanning the range of masses where the observed nuclear
recoil 
excesses might be consistent with exclusion limits from other 
experiments. We fix the elastic scattering cross-section on the proton to 
$\sigma_{\rm p} = 10^{-4} \, {\rm pb}$. It is straight-forward to scale 
our results to other values for the cross-section.
The number of events required to detect anisotropy, $N_{\rm iso}$, is 
independent of the cross-section, while the corresponding exposure, 
${\cal E}$, is calculated as
\begin{equation}
\label{exposeq}
{\cal E} = \frac{N_{\rm iso}}{\sigma_{\rm p} \, \rho \, R(>E_{\rm th}) } \,,\end{equation}
where $R(>E_{\rm th})$ is the total event rate (i.e. the integral of
the differential event rate) above the energy threshold normalised to
unit cross-section and local WIMP density. Therefore the
exposures required can be simply scaled for other cross-sections and
local WIMP densities.

\subsection{WIMP velocity distribution}
The detailed angular dependence of the recoil rate depends on the
exact form of the WIMP velocity distribution~\cite{copi:krauss,pap1,alenazi}. 
However, if the WIMP velocity distribution is dominated by a smooth 
component the main feature of the recoil distribution (namely the 
rear-front asymmetry) is robust (see e.g. Ref.~\cite{pap1})).
The number of events required to detect anisotropy depends relativity 
weakly on the WIMP speed distribution~\cite{pap1}. However the event rate 
above the energy threshold, and hence the exposure required to detect 
anisotropy, depends more significantly on the WIMP speed distribution.

Usually the dominant uncertainty in the (time and direction averaged)
differential event rate comes from the $\sim 10\%$ uncertainty~\cite{mb} 
in the value of the local circular speed, and hence the WIMP velocity 
dispersion (e.g. Refs.~\cite{br,greenfv,mmm,mccabe}). The velocity 
dispersion and WIMP mass have somewhat degenerate effects on the 
differential event rate. For instance if the velocity dispersion is 
increased, then there are more WIMPs with higher speeds, however the energy 
spectra of the resulting nuclear recoils can remain the same if the WIMP 
mass is decreased. Therefore the range of WIMP masses corresponding to a
particular observed energy spectrum excess moves to lower masses if
the velocity dispersion is increased (see e.g. Ref.~\cite{hk} for the
specific case of CoGeNT). Varying the velocity dispersion has a
qualitatively similar effect on the values of the WIMP mass consistent
with the DAMA annual modulation~\cite{savagevc}. Consequently, while 
varying the circular speed affects the values of the WIMP mass 
corresponding to, or excluded by, the various data sets, it does not 
significantly affect their compatibility. Therefore we fix the local 
circular speed to its standard value, $v_{\rm c} = 220 \, {\rm
  km \, s}^{-1}$, consistent with our benchmark WIMP masses 
  $m_{\chi} = 5, 7.5$ and $10 \, {\rm GeV}$.

Since the differential event rate involves an average over the WIMP
speed distribution it is usually relatively weakly sensitive to the
detailed shape of the speed distribution (e.g. Refs.~\cite{kk,greenfv}). 
This is not necessarily the case, however, for experiments that are only 
sensitive to the high speed tail of the distribution 
(e.g. Refs.~\cite{lsww,fpsv,mmm,mccabe}). The minimum WIMP speed which can 
cause a recoil of energy $E$, $v_{\rm min}$, is given by
\begin{equation}
v_{\rm min} = \left( \frac{E m_{\rm A}}{2 \mu_{\chi A}^2} \right)^{1/2} \,,
\end{equation}
where $m_{\rm A}$ is the mass of the target nuclei and $\mu_{\chi A}$
is the reduced mass of the WIMP-target nucleus system.

Particles with speed greater than the local escape speed, $v_{\rm esc}
\equiv \sqrt{2  |\Phi(R_{0})| } $ where $\Phi$ is the potential and
$R_{0}$ the Solar radius, will not be gravitationally bound to the
Milky Way~\footnote{Dark matter halos do contain unbound particles, 
however the fraction of such particles at the Solar radius is 
small~\cite{blw}.}. The RAVE survey found that the escape speed lies in 
the range $498 \, {\rm km \, s}^{-1} < v_{\rm esc} < 608 \, {\rm km \, s}^{-1}$
at $90\%$ confidence, with a median likelihood of $v_{\rm esc} =544 \,
{\rm km \, s}^{-1}$~\cite{rave}.   The maximum WIMP
speed in the lab frame is $v_{\rm esc} + v_{\rm lab}(t)$, where
$v_{\rm lab}(t)$ is the speed of the Earth with respect to the
Galactic rest frame.  
This is made up of three components: the motion of the Local Standard of 
Rest (LSR), ${\bf v}_{\rm LSR}=(0, v_{\rm c}, 0)$ in Galactic coordinates, 
the Sun's peculiar motion with respect to the LSR, ${\bf v}_{\odot}^{\rm p}(11.1, \, 12.2, 7.3) \, {\rm km \,s}^{-1}$~\cite{schoenrich}, and the 
Earth's orbit about the Sun, ${\bf v}_{\rm e}^{\rm orb}(t)$.  It has a 
maximum value at $t_{0} \approx 153$ days (on June 2nd) of
$v_{\rm lab}^{\rm max}=v_{\rm lab}(t_{0}) \approx 248 \, {\rm km} \, {\rm s}^{-1}$.

\begin{figure}
\includegraphics[width=8.5cm]{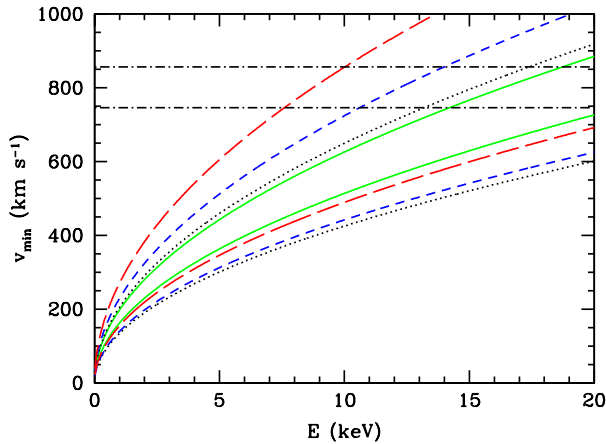}
\caption{The minimum WIMP speed, $v_{\rm min}$, which can cause a
recoil of energy $E$ for ${}^3{\rm He}$ (solid), C (dotted), F (short 
dashed) and S (long dashed) for (top and bottom curves in each case 
respectively) $m_{\chi}=5$ and $10 \, {\rm GeV}.$ The horizontal 
dot-dashed lines show the maximum WIMP speeds in the lab, 
$ v_{\chi}^{\rm max} \approx v_{\rm esc} +  248 {\rm km} \, {\rm s}^{-1}$, corresponding to the 90$\%$ upper and lower confidence limits on the 
escape speed from RAVE, $v_{\rm esc}^{\rm max}= 608 \, {\rm km} \, {\rm s}^{-1}$ and  $v_{\rm esc}^{\rm min}= 498 \, {\rm km} \, {\rm s}^{-1}$.  }
\label{vminfig}
\end{figure}

Fig.~\ref{vminfig} shows $v_{\rm min}$ as a function of $E$ for 
$m_{\chi}= 5$ and $10$ GeV for ${}^3{\rm He}$, ${\rm C}$, ${\rm S}$ and 
${\rm F}$ target nuclei. The horizontal lines show the maximum WIMP speeds 
in the lab, $v_{\chi}^{\rm max} = v_{\rm esc} + v_{\rm lab}^{\rm max}$, corresponding to the 
90$\%$ upper and lower confidence limits on the escape speed from RAVE, 
$v_{\rm esc}^{\rm max}= 608 \, {\rm km} \, {\rm s}^{-1}$ and  
$v_{\rm esc}^{\rm min}= 498 \, {\rm km} \, {\rm s}^{-1}$. For light WIMPs, 
unless the target nuclei are light and the threshold energy low, the 
minimum speed corresponding to the threshold energy lies in the tail of 
the speed distribution, and in some cases beyond the cut-off due to the 
Galactic escape speed. The expected event rates are therefore very 
sensitive to the value of the escape speed and the shape of the high 
speed tail of the distribution. 

The standard halo model,  an isotropic, isothermal sphere with
density profile $\rho(r) \propto r^{-2}$, is formally infinite. 
Hence its Maxwellian velocity distribution,
\begin{equation}
\label{max}
f({\bf v}) = N
 \exp{\left( - \frac{3|{\bf v}|^2}{2 \sigma^2} \right)}  \,, \\
\end{equation}
where $\sigma = \sqrt{3/2} v_{\rm c}$ and $N$ is a normalisation constant, 
extends to infinity too. This is usually addressed by truncating the 
velocity distribution by hand, either sharply or exponentially, at the 
escape speed.

Numerical simulations find velocity distributions with less high speed
particles than the standard Maxwellian~\cite{fairs,vogelsberger,kuhlen}. 
Lisanti et al.~\cite{lsww} have presented an ansatz for the velocity 
distribution which reproduces this behaviour: 
\begin{equation}
\label{k}
f(|{\bf v}|) \propto \left[ \exp{ \left( \frac{v_{\rm esc}^2 - |{\bf v}|^2}{k
        v_{0}^{2}} \right)} - 1 \right]^{k} \Theta( v_{\rm esc} -
|{\bf v}| ) \,.
\end{equation}
The parameter $k$ is related to the outer slope of the density profile,
$\gamma$, ($\rho(r) \propto r^{-\gamma}$ for large $r$), by
$k= \gamma- 3/2$ for $\gamma > 3$~\cite{kochanek}.

\begin{figure}
\includegraphics[width=8.5cm]{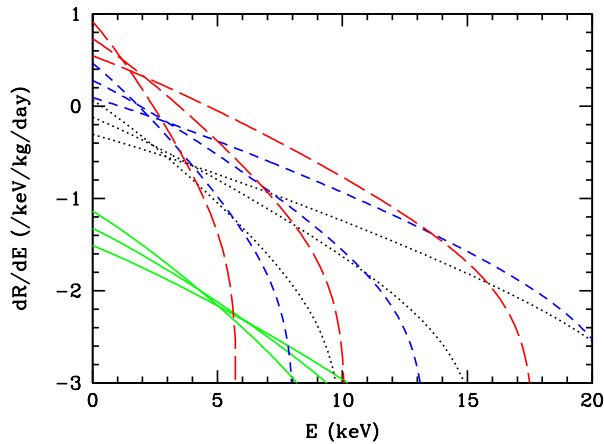}
\caption{The differential event rates, assuming a Maxwellian speed distribution with a sharp cut-off at $v_{\rm  esc}^{\rm max}= 608 \, {\rm km \, s}^{-1}$, a cross-section $\sigma_{\rm p} =
  10^{-4} \, {\rm pb}$ and a local WIMP density $\rho= 0.3 \, {\rm
    GeV} \, {\rm cm}^{-3}$, for He (solid), C (dotted), F (short
  dashed) and S (long dashed) for (from top to bottom at $E=0 \, {\rm
    keV}$) $m_{\chi}= 5, 7.5$ and $10 \, {\rm GeV}.$ 
 }
\label{drde}
\end{figure}

We consider 2 forms for $f({\bf v})$ chosen to have high speed tails
which roughly span the plausible range. 
Firstly, a Maxwellian distribution,  
eq.~(\ref{max}),  with a sharp cut-off ($f({\bf v}) = 0$ for $|{\bf v}| \geq v_{\rm esc}^{\rm max}$)   at the upper limit on the escape speed from 
RAVE, $v_{\rm esc}^{\rm max}= 608 \, {\rm km \, s}^{-1}$, which 
provides a large tail event rate. 
Secondly, a Lisanti et al. $f(v)$, eq.~(\ref{k}), with 
$v_{\rm esc}= v_{\rm esc}^{\rm min}= 498 \,{\rm km \, s}^{-1}$ 
to provide a small tail event rate. 
We fix $k=1.5$, corresponding to an outer density 
profile slope $\gamma=3$ and $v_{0}= v_{\rm c} = 220 \,{\rm km \, s}^{-1}$.
In both cases we fix the local density to be 
$\rho=0.3 \, {\rm GeV} \, {\rm cm}^{-3}$. 
Scaling the event rates and exposures to other densities is 
straight-forward.

Fig.~\ref{drde} shows the differential event rates for each of the WIMP 
masses and targets we consider calculated using the Maxwellian 
distribution with a sharp cut-off at $v_{\rm  esc}^{\rm max}= 608 \, {\rm km \, s}^{-1}$. 
The much smaller event rates for ${\rm He}$ are largely due to the $A^2$ 
factor in the event rate for spin-independent scattering. For heavier 
targets the differential event rate in the $E \rightarrow 0$ limit is 
substantially larger, however it decreases rapidly with increasing $E$, 
and the lighter the WIMP the more rapid the decrease.  Furthermore for 
relatively small energies $v_{\rm min}$ exceeds the maximum WIMP speed in 
the lab frame and hence the differential event rate is zero.

\begin{figure}
\includegraphics[width=8.5cm]{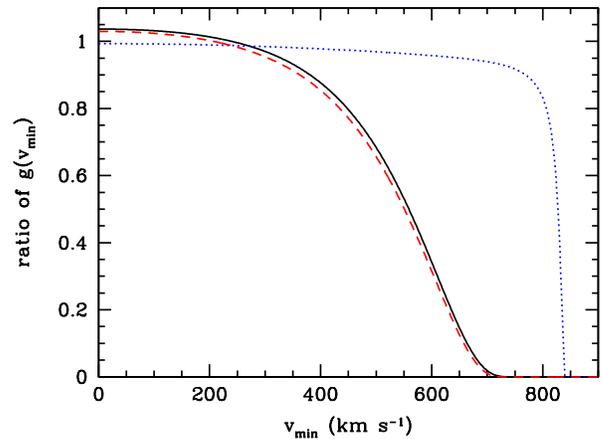}
\caption{The speed integral, $g(v_{\rm min})$, relative to that for the Maxwellian distribution with a sharp cut-off at $v_{\rm  esc}^{\rm max}= 608 \, {\rm km \, s}^{-1}$, including averaging over the Earth's orbit. The solid line is for Lisanti et al.'s $f(v)$, eq.~(\ref{k}), with $v_{\rm esc}^{\rm min}= 498 \,
{\rm km \, s}^{-1}$, including the Earth's orbit. The dotted and dashed lines are for the Maxwellian and  Lisanti et al. distributions respectively,  neglecting the Earth's orbit. }
\label{ratio}
\end{figure}

Fig.~\ref{ratio} shows the ratio of the speed integral, 
\begin{equation}
g(v_{\rm min}) = \int_{v_{\rm min}}^{\infty} \frac{f(v)}{v} \,{\rm d} v \,, 
\end{equation}
for Lisanti et al.'s $f(v)$ in eq.~(\ref{k}) with $v_{\rm esc}^{\rm min}= 498 \,
{\rm km \, s}^{-1}$ to that for the Maxwellian distribution with a sharp 
cut-off at $v_{\rm  esc}^{\rm max}= 608 \, {\rm km \, s}^{-1}$. It also 
shows the velocity integral ratios for these two models if the Earth's 
orbit is neglected.  For $v_{\rm min} \lesssim {\cal O}(300 \, {\rm km} \, {\rm s}^{-1})$ the difference in the speed integrals for the two speed 
distributions is relatively small, less than $10\%$. As $v_{\rm min}$ is 
increased, so that only the tail of the speed distribution is included in 
the speed integral, the difference becomes substantially larger, reaching 
roughly an order of magnitude for $v_{\rm min} \sim  650  \, {\rm km} \, {\rm s}^{-1}$.  
Neglecting the Earth's orbit has a significant ($>10\%$) effect, for 
$v_{\rm min}$ within $\sim 100  \, {\rm km} \, {\rm s}^{-1}$ of the 
maximum WIMP speed in the lab frame $v_{\chi}^{\rm max}$. Therefore the 
Earth's orbital speed must be included, and averaged over, for an accurate 
calculation of the event rate.

Simulated dark matter halos have velocity distributions which contain
features at high speeds~\cite{vogelsberger,kuhlen}. More specifically
there are fairly broad features, which are similar at different
positions within a single halo, but which vary from halo to halo, and
are hence thought to be a relic of the formation history of the
halo~\cite{vogelsberger,kuhlen,ls}. Ref.~\cite{kuhlen} also finds narrow
features in some locations, corresponding to tidal streams, while 
Ref.~\cite{purcell} finds that the 
dark matter streams from the Sagittarius dwarf are significantly more 
extended than the stellar streams, and the leading dark matter stream may 
pass through the Solar neighbourhood.  The
detailed shape of the high speed tail of $f(v)$ would affect the
interpretation of the CoGeNT, CRESST and DAMA data, in particular the
values of the WIMP mass extracted. Since no detailed study of
this has been carried out to date~\footnote{Ref.~\cite{khb} includes tidal 
streams with specific properties chosen to reproduce the energy dependence
of the amplitude of the annual modulation measured by CoGeNT, while
Ref.~\cite{nsf} studies the effects of the Sagittarius tidal stream.}, 
we do not include the high speed features in our analysis.  We defer a
general investigation of the directional event rate produced by
simulation velocity distributions to future work~\cite{inprep}.

\subsection{Statistical tests}
We follow the statistical procedures described in detail in
Refs.~\cite{pap1,pap3}.  We use the Rayleigh-Watson statistic, which
uses the mean resultant length of the recoil direction vectors. We
also use the Bingham statistic which, unlike the Rayleigh-Watson
statistic, can be used with axial data (where the senses of the nuclear 
recoils are not measured). These statistics are described in more detail in
Appendices \ref{rwstat} and \ref{bingstat}.

For each WIMP mass, target nuclei and energy threshold combination we 
calculate the probability distribution of each statistic for WIMP induced 
recoils and also for the null hypothesis of isotropic recoils. We use 
these distributions to calculate the rejection and acceptance factors, 
$R$ and $A$. The rejection factor gives the confidence level with which
the null hypothesis can be rejected given a particular value of the
test statistic, while the acceptance factor is the probability of
measuring a larger value of the test statistic if the alternative
hypothesis is true. We then find the number of events required for 
$A = R = 0.95, 0.997 \, 30$ and $0.999\, 999 \,427$ i.e. to reject isotropy at
this confidence level in this percentage of experiments (see
Refs.~\cite{pap1,pap3} for further discussion). The later two
confidence levels correspond, for a gaussian distribution, to three
and five sigma respectively.  Since our aim is to examine whether 
directional detection experiments could detect light WIMPs we will focus 
on the case $A = R = 0.999\, 999 \,427$, corresponding to the `five sigma' 
result conventionally  required for discovery.

\section{Results and discussion}
\label{res}
For each of the combinations of WIMP mass, target nuclei, energy
threshold and confidence level discussed in Sec.~\ref{model} we
calculate the number of events required to reject isotropy, $N_{\rm iso}$, 
for vector and axial data (using the Rayleigh-Watson and
Bingham statistics respectively). The number of events required for a
$5\sigma$ detection with the Rayleigh-Watson statistic varies
from 7 to 58. For fixed WIMP and target mass, $N_{{\rm iso}}$ decreases 
with increasing energy threshold, $E_{\rm th}$, (since the recoils caused 
by high speed WIMPs in the tail of the distribution are more anisotropic), 
until the minimum speed required to cause a recoil of energy $E_{\rm th}$, 
$v_{\rm min}(E_{\rm th})$, exceeds the maximum WIMP speed in the lab 
frame, $v_{\chi}^{\rm max}$. At this point the event rate is zero and 
no events can be detected. The same trend occurs for decreasing WIMP mass 
(with threshold energy and target mass fixed). For fixed threshold 
energy, $N_{{\rm iso}}$ decreases with increasing target mass for 
$m_{\chi}= 5 \, {\rm GeV}$, however for $m_{\chi}=7.5$ and 
$10 \, {\rm GeV}$, $N_{{\rm iso}}$ increases as the target mass number is 
increased from $A=3$ to $12$ before decreasing as $A$ is increased further. 
This is due to the variation of $v_{\rm min}$ with target and WIMP mass 
shown in Fig.~\ref{vminfig}.  

For the Bingham statistic, which can be used with axial data, the number 
of events required for a $5\sigma$  detection varies from 9 to more than 
190~\footnote{In a small number of cases, where a large number of events 
are required, we have only been able to place a lower limit on 
$N_{\rm iso}$ due to computational time limitations.}. Both the number of 
events and its increase, relative to the number required for the 
Rayleigh-Watson statistic, is smallest for the configurations which are 
only sensitive to the highly anisotropic recoils from high speed WIMPs in 
the tail of the speed distribution.

\begin{figure}[t]
\includegraphics[width=8.5cm]{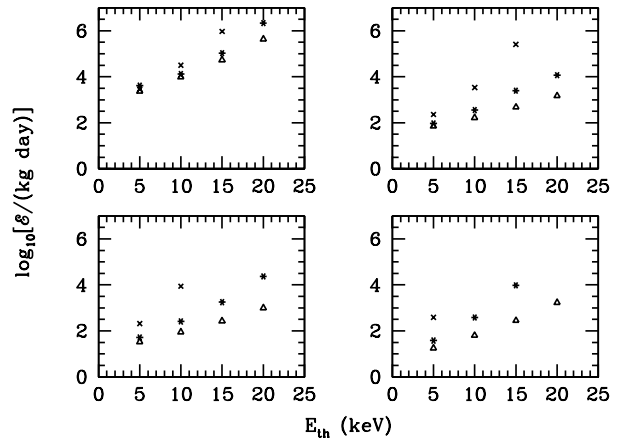}
\caption{The exposure required for a $5 \sigma$ rejection of isotropy 
  as a function of energy threshold, $E_{\rm th}$, for  ${}^3{\rm He}$ (top left), C
  (top right), F (bottom left) and S (bottom right) for $m_{\chi}= 5,
  7.5$ and $10 \, {\rm GeV}$ (crosses, stars and triangles respectively) for the 
  Maxwellian $f(v)$ with a sharp cut-off at $v_{\rm  esc}^{\rm max}= 608 \, {\rm km \, s}^{-1}$.
 Where a symbol is not displayed, the minimum WIMP speed corresponding to the energy threshold,
 $v_{\rm min}(E_{\rm th})$, for this WIMP and target mass combination exceeds the maximum WIMP speed in the lab frame, $v_{\chi}^{\rm max}$ and hence the event rate is zero.}
\label{expostargetRW5sigMax}
\end{figure}

If an experiment is only sensitive to high speed WIMPs, fewer events are 
required to reject isotropy, how- ever, the reduced event rate means that 
the exposure required to accumulate these events will be larger.
We therefore use eq.~(\ref{exposeq}) to calculate the exposure, ${\cal E}$,
(in ${\rm kg \, day}$) required to accumulate the required number of events
for each case. As illustrated in Fig.~\ref{drde}, for light WIMPs the
differential event rate decreases rapidly with increasing energy, and
therefore the event rate above the energy threshold, $R(>E_{\rm th})$,
plays a crucial role in determining the exposure required.

The exposure required to reject isotropy at $5 \sigma$ using the
Rayleigh-Watson statistic assuming a Maxwellian $f(v)$ with a sharp 
cut-off at $v_{\rm  esc}^{\rm max}= 608 \, {\rm km \, s}^{-1}$ is shown 
for each configuration in
Fig.~\ref{expostargetRW5sigMax}. While $N_{{\rm iso}}$ varies by less than 
an order of magnitude, because of the large
variation in $R(>E_{\rm th})$, the exposure varies
by more than five orders of magnitude.
Due to the rapid decrease of $R(>E_{\rm th})$, the exposure increases
sharply with increasing $E_{\rm  th}$ for each WIMP and target nuclei mass
combination. The factor by which the exposure increases, increases
with both decreasing WIMP mass and increasing target nuclei mass (i.e. as
the minimum WIMP speed to which the experiment is sensitive is increased).
Eventually the minimum WIMP speed corresponding to the energy threshold, 
$v_{\rm min}(E_{\rm th})$,  exceeds the maximum WIMP speed in the lab 
frame, $v_{\chi}^{\rm max}$, and the event rate is zero and WIMPs of 
this mass can not be detected.

\begin{figure}
\includegraphics[width=8.5cm]{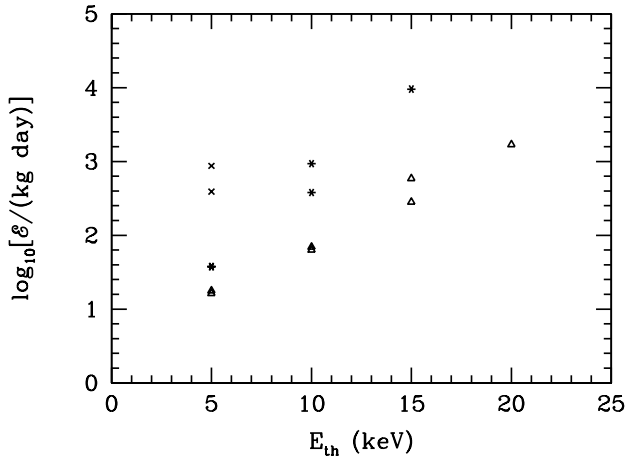}
\caption{As Fig.~\ref{expostargetRW5sigMax} but for a S target only comparing  the exposures required
for the  Lisanti et al. $f(v)$, eq.~(\ref{k}), with $v_{\rm esc}^{\rm min}= 498 \,
{\rm km \, s}^{-1}$ (upper symbols) with those for the Maxwellian $f(v)$ with a sharp cut-off at $v_{\rm  esc}^{\rm max}= 608 \, {\rm km \, s}^{-1}$ (lower symbols). For $m_{\chi} = 7.5 \, {\rm keV}$ and $E_{\rm th}= 5 \, {\rm keV}$ the exposures required for the two speed distributions are the same, and hence only one symbol is visible.}
\label{expostargetRW5sigS}
\end{figure}

Of the halo models considered, the Maxwellian $f(v)$ with a sharp cut-off 
at $v_{\rm  esc}^{\rm max}= 608 \, {\rm km \, s}^{-1}$ 
has the largest tail event rate, and hence the smallest exposures. 
In Fig.~\ref{expostargetRW5sigS} we show the exposures for a ${\rm S}$ 
target for the Lisanti et al. $f(v)$, eq.~(\ref{k}), with 
$v_{\rm esc}^{\rm min}= 498 \,{\rm km \, s}^{-1}$ as well. 
When $v_{\rm min}(E_{\rm th})$ is much smaller than $v_{\chi}^{\rm max}$
the exposure required is fairly modest, 
${\cal E} \sim 10-100 \, {\rm kg} \, {\rm day}$ and the event rates, and 
hence exposures, for the two speed distributions are very similar. However 
as $v_{\rm min}(E_{\rm th})$ approaches $v_{\chi}^{\rm max}$ the 
exposures required become large and the differences between the two speed 
distributions become significant. For instance for 
$E_{\rm th} = 20 \, {\rm keV}$, WIMPs with $m_{\chi} =10 \, {\rm GeV}$ 
and a Maxwellian distribution with  
$v_{\rm  esc}^{\rm max}= 608 \, {\rm km \, s}^{-1}$ anisotropy could be 
detected with an exposure of $1700  \, {\rm kg} \, {\rm day}$, however if 
the WIMPs have the Lisanti et al. $f(v)$ with 
$v_{\rm esc}^{\rm min}= 498 \, {\rm km \, s}^{-1}$ the event rate is zero 
and they can not be detected. The trends for the other target nuclei are 
similar.

\begin{figure}
\includegraphics[width=8.5cm]{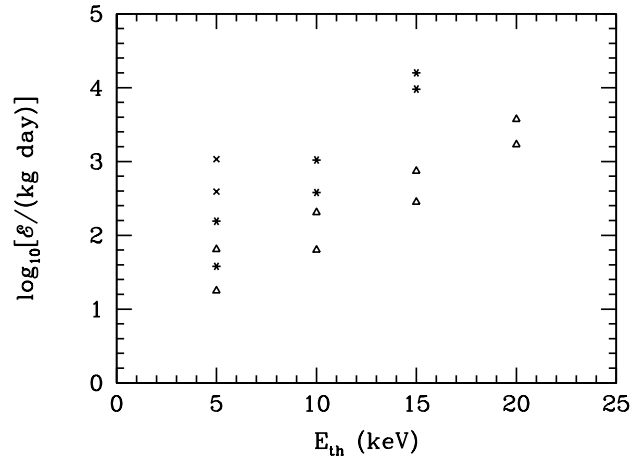}
\caption{As Fig.~\ref{expostargetRW5sigS} but comparing  the exposures required using the Bingham statistic (upper symbols) with those required using the Rayleigh statistic (lower symbols), in both cases for a S target using a Maxwellian $f(v)$ with a sharp cut-off at $v_{\rm  esc}^{\rm max}= 608 \, {\rm km \, s}^{-1}$.
For $m_{\chi} = 10 \, {\rm GeV}$ and $E_{\rm th}=5 \, {\rm keV}$ we have only been able to place a lower bound on the number of events, and hence exposure, required with the Bingham statistic. }
\label{exposRWB5sigS}
\end{figure}

In Fig.~\ref{exposRWB5sigS} we compare the exposures required to reject 
isotropy with axial data using the Bingham statistic with those for 
vectorial data using the Rayleigh-Watson statistic, for a ${\rm S}$ target 
and a Maxwellian $f(v)$ with a sharp cut-off at 
$v_{\rm  esc}^{\rm max}= 608 \, {\rm km \, s}^{-1}$. 
Since $R(>E_{\rm th})$ for each configuration doesn't change, the 
variations in the exposure are driven entirely by the variations in 
$N_{\rm iso}$ discussed above. Therefore the increase in the exposure, 
relative to that required for the Rayleigh-Watson statistic, is smallest 
for the cases which are only sensitive to the highly anisotropic recoils 
from high speed WIMPs in the tail of the speed distribution. However in 
these cases the exposure is large even for the Rayleigh-Watson statistic, 
due to the small values of $R(>E_{\rm th})$. 

We have focused on the number of events and exposure required for 
a `$5 \sigma$' discovery of light WIMPs with a directional detector. 
Significant experimental support for light WIMPs would be obtained even 
with a lower significance signal. For $95\%$ and $99.713\%$ confidence 
levels (the later corresponding to $3 \sigma$) the number of events, and 
hence the exposure, required with the Rayleigh-Watson statistic is smaller 
by a factor of between $0.40-0.91$ and $0.69-0.95$ respectively. The 
factor by which the number of events must be increased to increase the 
significance of the rejection of isotropy is
smallest when the experiment is only sensitive to the most anisotropic
events coming from the tail of the speed distribution. However in these 
cases the exposure required even for a low significance detection is large.

The final question is `How achievable are these exposures and energy thresholds by current and 
near-future detectors'?  A typical current detector consisting of a 
$ 1 \, {\rm m}^{3} $ TPC filled to 75 Torr with 
${\rm CF}_{4}$ or ${\rm  CS}_{2}$ could, in roughly a year, achieve an 
exposure of order $10^{3} \, {\rm kg \, day}$~\cite{cygnus}. 
A $10^{3} \, {\rm kg \, day}$ exposure with ${\rm CF}_{4}$ or 
${\rm  CS}_{2}$ would (provided that recoils are measured in 3d with 
good, $\lesssim 10^{\circ}$, angular resolution) be capable of 
detecting $m_{\chi}=10 \, {\rm keV}$ WIMPs with an energy threshold of 
$20 \, {\rm keV}$ or lower. Lighter WIMPs would require a lower
energy threshold, potentially as low as $5 \,{\rm keV}$ for 
$m_{\chi}= 5 \, {\rm GeV}$. With a ${}^3{\rm He}$ target a 
low, $\sim 5 \, {\rm keV}$, energy threshold would be required, even 
for $m_{\chi}= 10 \, {\rm GeV}$. 
Measuring the directions of low energy nuclear recoils is a major experimental challenge, and
these energy threshold are lower than those which have been used in the analysis of data from current, prototype, detectors~\cite{cygnus}. One of the focuses of the R\&D for future generation experiments is to reduce the energy threshold (see e.g. \cite{cygtalks}).
The MIMAC experiment has, using micromegas readout, detected 
5 keV F recoils~\cite{mimaclow}, while DRIFT-II is sensitive to nuclear recoils down to sub $5$ keV energies~\cite{driftlow}.  Simulations of the MIMAC detector indicate that the directional energy threshold will lie below $20$ keV~\cite{billardtrack}

\section{Summary}
\label{dis}
The event rate excess and annual modulations observed by various direct 
detection experiments may be due to light, $5-10 \, {\rm GeV}$ WIMPs. We 
have investigated whether near future directional detection experiments 
will be able to test this possibility, by detecting the anisotropy of the 
nuclear recoils. We find, using the Rayleigh-Watson statistic, that an 
ideal directional detector (capable of measuring the directions of the 
recoils, including their senses, in 3d with good angular resolution) would 
require between 7 and 58 events to detect the anisotropy at $5 \sigma$. 
The number of events required depends on the target nuclei mass, energy 
threshold, WIMP mass, and crucially the details of the high speed tail of
the WIMP speed distribution. It is smallest for cases where the 
experiment is only sensitive to the highly anisotropic recoils from high 
speed WIMPs in the tail of the speed distribution. If the detector is not 
capable of measuring the senses of the recoils we find, using the Bingham 
statistic, that the number of events required ranges between 9 and more 
than 190. The increase in the number of events required with axial data is 
smallest for the cases which are only sensitive to high speed WIMPs due
to the higher degree of anisotropy in the nuclear recoil distributions.

In terms of the detection potential the key quantity is the exposure 
necessary to detect the required number of events, which is inversely 
proportional to the event rate above threshold. For the configurations 
we have considered (${}^3{\rm He}$, C, F and S targets, 
$5-10 \, {\rm GeV}$ WIMPs and energy threshold between $5$ and 
$20 \, {\rm keV}$) the event rate above threshold varies by more than 
five orders of magnitude, and in some cases the minimum speed required to 
cause a recoil above threshold exceeds the maximum WIMP speed in the lab 
and the event rate is zero. For the cases which are only sensitive to 
high speed WIMPs, where the number of events required was smallest, the 
event rate above threshold is small and hence very large exposures would 
be required.  The shape of the high speed tail of the WIMP distribution 
is not well known and this leads to large uncertainties in the event rate 
expected in experiments which are only sensitive to the high speed tail, 
c.f. Refs.~\cite{lsww,fpsv,mmm,mccabe}. We also emphasize that including, 
and averaging over, the Earth's orbit is essential for an accurate 
calculation of the event rate in these cases.

We find that a future ${\rm CF}_{4}$ or ${\rm  CS}_{2}$ detector with an energy 
threshold of $20 \, {\rm keV}$ or lower, which can measure recoil 
directions and senses in 3d with good angular resolution, would be
capable of detecting WIMPs with $(m_{\chi}, \sigma_p) = (10 \, {\rm GeV},
10^{-4} \, {\rm pb})$ with an exposure of $10^{3} \, {\rm kg \, day}$.
Detecting lighter WIMPs would require a lower energy threshold. With a 
${}^3{\rm He}$ target a low, $\sim 5 \, {\rm keV}$, energy threshold would 
be required, even for WIMP masses at the upper end of the mass range 
considered. 
In summary, we conclude that directional detection experiments may be 
able to detect light WIMPs, but this depends quite sensitively on 
both the experimental configuration (target nuclei mass and energy 
threshold) and the unknown WIMP mass and velocity 
distribution.  The directional energy thresholds required to detect light WIMPs are below those used in analyses of data from 
current directional detectors, however it has been demonstrated that directional detectors can detect~\cite{mimaclow,driftlow} and measure the directions~\cite{billardtrack} of low energy nuclear recoils and R\&D is underway to realise lower energy thresholds~\cite{cygtalks}.

We also note 
that directional experiments, with heavy targets, could also test 
inelastic dark matter as the explanation of the direct detection 
anomalies~\cite{inelastic}. 

\begin{acknowledgments}
  AMG and BM are supported by STFC. 
 \end{acknowledgments}

\appendix

\section{Rayleigh Watson statistic}
\label{rwstat}

The (modified) Rayleigh-Watson statistic, ${\cal W}^{\star}$ is the simplest coordinate independent
statistic for detecting anisotropy in vectorial data. It is related to
the Rayleigh statistic, ${\cal R}$, which for a sample of $N$ unit
vectors $\vec{x}_i $ is given by
\begin{equation}
{\cal R} = \left | \sum^N_{i=1} \vec{x}_i \right | \,.
\end{equation}
i.e. the modulus of the sum of vectors. For an isotropic data set
${\cal R}$ should be zero, modulo statistical fluctuations. For
anisotropic data the value of ${\cal R}$ becomes larger as the degree of anisotropy increases.

The modified Rayleigh-Watson statistic, ${\cal W}^{\star}$, defined as~\cite{watson1,watsonbook,mardia:jupp}
\begin{equation} 
{\cal W}^{\star} = \left( 1- \frac{1}{2N} \right) {\cal W} +
\frac{1}{10N} {\cal W}^2 \,,
\end{equation}
where ${\cal W}$ is the (unmodified)  Rayleigh-Watson statistic 
\begin{equation}
{\cal W} = \frac{3}{N}  {\cal R}^2 \,.
\end{equation}
The modified statistic ${\cal W}^{\star}$ has the advantage of
approaching its large $N$ asymptotic distribution for smaller $N$ than
the unmodified statistic. For isotropically distributed vectors,
${\cal W}^{\star}$ is asymptotically distributed as
$\chi^2_3$~\cite{watson1,watsonbook}.  The difference between
$\chi^2_3$ and the true distribution of ${\cal W}^{\star}$ for
isotropic vectors in the large ${\cal W}^{\star}$ tail of the
distribution is less than $2\%$ for $N>30$~\cite{pap1}. For smaller
$N$ the $\chi^2_3$ distribution significantly underestimates the true
probability distribution and therefore, as in Ref.~\cite{pap1} we
calculate the probability distribution from the exact probability
distribution of ${\cal R}$, as described in
Ref.~\cite{stephens:rayleigh}.

\section{Bingham statistic} 
\label{bingstat}
The Rayleigh-Watson statistic can not be used with axial data, as it
is not sensitive to distributions which are symmetric with respect to
the centre of the sphere. For axial data the Bingham statistic ${\cal B}^{\star}$
which is based on the scatter matrix of the data, ${\bf T}$, can be
used. This matrix is defined as~\cite{watson2,watsonbook,mardia:jupp}
\begin{equation}
{\bf T} = \frac{1}{N}
\sum^N_{i=1}
\left( \begin{array}{ccc}
x_i x_i & x_i y_i & x_i z_i \\
y_i x_i & y_i y_i & y_i z_i \\
z_i x_i & z_i y_i & z_i z_i
\end{array} \right) \,,
\end{equation}
where $(x_i,y_i,z_i)$ are the components of the $i$-th vector or axis.
This matrix is real and symmetric with unit trace, so that that the
sum of its eigenvalues $e_k$ ($k=1,2,3$) is unity, and for an
isotropic distribution all three eigenvalues should, modulo
statistical fluctuations, be equal to $1/3$.  The Bingham 
statistic, ${\cal B}$, 
\begin{equation}
{\cal B} = \frac{15N}{2}\sum^3_{k=1}\left( e_k - \frac{1}{3}
\right)^2 \,,
\end{equation}
measures the deviation of the
eigenvalues $e_k$ from the value of 1/3 expected for an isotropic
distribution.  For isotropically distributed vectors/axes ${\cal
B}$ is asymptotically distributed as $\chi^2_5$. Since ${\bf T}$ is symmetric
under a sign change of $\vec{x}$, the Bingham statistic can be used for
axial data as well as vectors.


\end{document}